# Self-Organization in Traffic Lights: Evolution of Signal Control with Advances in Sensors and Communications

Sanjay Goel[1], Stephen F. Bush[2], Carlos Gershenson[3]

[1] School of Business, University at Albany, SUNY
[2] GE Global Research
[3] Instituto de Investigaciones en Matemáticas Aplicadas y en Sistemas & Centro de Ciencias de la Complejidad, Universidad Nacional Autónoma de México; SENSEable City Lab, Massachusetts Institute of Technology; ITMO University

**Abstract**—Traffic signals are ubiquitous devices that first appeared in 1868. Recent advances in information and communications technology (ICT) have led to unprecedented improvements in such areas as mobile handheld devices (*i.e.*, smartphones), the electric power industry (*i.e.*, smart grids), transportation infrastructure, and vehicle area networks. Given the trend towards interconnectivity, it is only a matter of time before vehicles communicate with one another and with infrastructure. In fact, several pilots of such vehicle-to-vehicle and vehicle-to-infrastructure (e.g. traffic lights and parking spaces) communication systems are already operational. This survey of autonomous and self-organized traffic signaling control has been undertaken with these potential developments in mind. Our research results indicate that, while many sophisticated techniques have attempted to improve the scheduling of traffic signal control, either real-time sensing of traffic patterns or a priori knowledge of traffic flow is required to optimize traffic. Once this is achieved, communication between traffic signals will serve to vastly improve overall traffic efficiency.

## I. INTRODUCTION

Traffic congestion is a perpetual source of frustration, lost productivity and greenhouse gas emissions. In 2011, U.S. commuters are estimated to have wasted 2.9 billion gallons of fuel, equivalent to $121 billion, stuck in traffic [1]. Traffic signals regulate the flow of traffic through intersections and improve throughput. However, if configured incorrectly, they can have a net negative impact on traffic flow. Consequently, optimization of traffic signals is essential to the operation of the traffic grid. This is a complex multivariate optimization problem where road network topology, intersection density, red signal probabilities, distance, speed, traffic density and other variables have to be considered.

Moreover, traffic light coordination is an EXP-complete problem [2], so large networks cannot be optimized in practice. Since the precise traffic demand is changing constantly, any optimal scheme soon becomes obsolete.

Traffic control measures have long sought to address these issues and there have been many improvements made over the course of the last century. More recently, however, these improvements have plateaued. Although research on traffic control systems continues, as this review suggests, according to Zhao and Tian [3] most deployments of adaptive traffic control systems in the U.S., for example, consist of the SCOOT or SCATS systems developed in the 1970s. In this paper we show that the concept of *self-organization* can leverage the advances in sensors and communication to provide a new leap in traffic light coordination, as traffic lights can systematically adapt to changing demands. This paper first provides a historical perspective on the development of a variety of traffic control systems and techniques, leading to a review of *self-organized traffic control*. We provide a discussion of the basic components and characteristics of self-organized traffic control systems, and their importance for future traffic optimization.

*1.1 The Evolution of Traffic Control Systems*

The first traffic signal was installed outside the British Houses of Parliament by J. P. Knight in 1868. The signal consisted of a gas lantern that exploded a year later, severely injuring the operator [4]. The first electric signal, designed by Lester Wire [5], was installed in Cleveland, Ohio in 1912 [6]. By 1917, interconnected traffic signal systems were in use in Salt Lake City. In 1922, the first automatically controlled interconnected system of traffic signals was installed in Houston, Texas [7].

Many traffic signals still function at a single intersection, and have pre-stored signal timing plans based on historically determined flow patterns. Initial attempts to improve these signals were made by applying Erlang's (1909) queuing theory, and F.V. Webster's (1958) queuing model for determining the optimal traffic signal cycle time. Refinements to Webster's model continue to be made [8]; [9]; [10]; [11]. Since their initial usage, several attempts have been made to create a closed-form solution to address the drawbacks of this method [12]. Such systems are simple, but despite all attempts to optimize them, they remain inefficient primarily because they are not based on real-time conditions. Since traffic demand changes with each cycle, so too do the number of queuing vehicles and the idling green times, and thus the optimal phases. Manual programming of each signal also makes frequent adjustments laborious and cumbersome. To regulate traffic at intersections, roundabouts are efficient for low speed, low-density roads, but not for high speed or high-density arteries [13] [14].

Modern traffic signals leverage sensors to adapt to prevalent conditions at the intersections. Such systems that respond to current traffic conditions are known as *adaptive traffic control systems* (ATCS). ATCS make signal timing decisions at short intervals based on real-time traffic conditions. These systems have been called third-generation systems [3]. ATCS still leave a lot of room for efficiency improvement; by leveraging sensors and communication traffic flow can be predicted better reducing wait times further. ATCS contrast with first-generation systems, which use historical data about traffic patterns to determine signal timing plans. ATCS are also an improvement to second-generation systems, which make use of signal timing plans based on surveillance data, but are restricted as to how frequently the response is generated [15]. As summarized in Table 1, we foresee self-organized systems as fourth-generation traffic control systems, providing a logical progression of improvement.

There are still a number of shortcomings with ATCS. Although ATCS are actuated signals, there is latency associated with the response of the signal to the trigger, causing inefficiencies. Moreover, adaptive signals work well for peak-traffic conditions; however, they can cause problems, for instance, when an entire lane of fast-moving traffic comes to a halt at a red signal that was triggered by a single vehicle arriving at a side road. These signals are expensive, fragile, and require constant maintenance. These and related problems are discussed in the authoritative traffic control handbook prepared for the U.S. Federal Highway Administration [16] .

**Table 1.** System types and their evolution

| Generation | System Type | Description |
| --- | --- | --- |
| *First-generation* | Pre-stored signal timing plans | Uses historical data about traffic patterns to determine signal timing plans. |
| *Second-generation* | Fixed-time signal timing plans based on surveillance data | Signal timing reacts to traffic conditions, but at fixed time intervals, and not in direct response to real-time traffic flows. |
| *Third-Generation* | Adaptive signals with synchronized and centralized or decentralized control | Signal timing reacts to real-time traffic conditions, making constant changes based on current traffic patterns. System is operated under centralized or decentralized control. |
| *Fourth-Generation* | Adaptive signals with decentralized, and self-organized control | Signal timing reacts to real-time traffic conditions at multiple intersections in a network through decentralized and distributed control. |

Attempts have been made to synchronize multiple signals together rather than manage them independently. This involves offsetting the timing of subsequent signals, such that a car traveling through intersections continues to see green light across multiple signals. This is usually done on the basis of traffic volume in one direction, and can often lead to unnecessary delay in the other direction. Several dynamic systems have been proposed for control based on traffic flows across multiple signals; however, most of these communicate with, and are controlled from, a central location. Issues associated with central control include scalability, resilience, and latency. As the number of traffic signals increases, so does the number of complicating variables; as the complexity increases exponentially, centrally controlled systems do not cope well with anomalies [2]. A variety of approaches involving centralized control have been used to optimize signal timings, such as: mixed-integer linear programming [17], [18], [19]; neural networks [20]; [21]; genetic reinforcement learning [22]; and fuzzy logic [23]; [24]. He et al. [25] use mixed integer programming to prioritize conflicting priority access to intersections of pedestrians and emergency vehicles, while optimizing signal coordination.

A model with independent controllers at each grid point that transmit data to a central controller for optimization and receive feedback in real-time is proposed by Li and colleagues [26]. This model sacrifices latency for computational power of a central control processor. Unfortunately, these models become more complex as the number of signals increases. Traffic control could be enhanced by analyzing static and mobile vehicle sensor data, although this would be extremely time-consuming. Other approaches to the problem involve graph theory and Petri nets [27]; [28], heuristic algorithms [29] [30] and meta-heuristic search algorithms [31]; [32]; [33]; [34]; [35]. A schematic illustrating different techniques and methods of traffic control is presented in Figure 1.

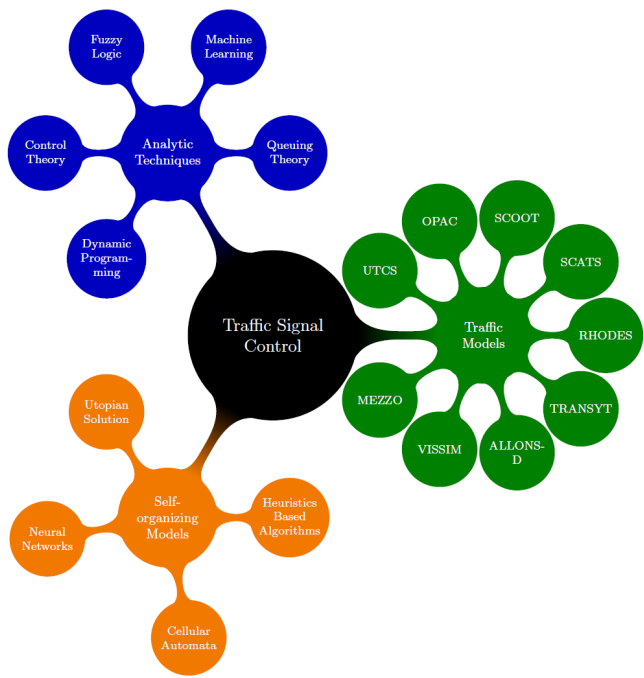

**Figure 1.** Traffic Signal Control Literature

*1.2 Argument and Outline*

Several innovative techniques can be applied to further improve traffic optimization, including queuing theory and graph spectra. For instance, the time that a car is "blocked" is determined by the number of red signals encountered en route to its destination; this can be determined analytically based on the density of intersections (the graph structure), the probability of red signals (the control algorithm), as well as the distance, speed, and density of traffic. This again becomes increasingly complex as the number of signals in the graph increases.

Ultimately, optimization of traffic systems, as in other mathematical applications, involves the use of an objective function to determine the best set of inputs from a set of available alternatives, which maximizes (or minimizes) some measure. In traffic studies, this measure usually includes average delays, stops or throughput. Although many of the methods and models discussed here may provide optimum solutions within a defined space, as traffic is non-stationary, the optimum changes constantly. Therefore, approaches based on objective functions do not always find the best possible traffic performance [36].

Decentralized, distributed, adaptive, and machine learning techniques for traffic signal control have been used with varying degrees of success. However, with recent advances in sensing and communication technologies, algorithms and techniques that were not previously thought applicable may now serve to improve traffic signal optimization. Self-organization is very effectively used in nature to manage complex systems with a large number of components [37]. Similarly, we believe that a distributed solution based on the concept of self-organization is possible in traffic control, and could provide the best means to optimize traffic flow patterns.

The outline of this paper is as follows. Section II reviews adaptive traffic control models that have been utilized in practice, as well as a number of traffic control models provided in the literature. We also review some aspects of urban traffic models. Section III examines in more depth particular techniques employed as part of traffic control systems. Section IV delves into self-organizing systems as an adaptive and efficient approach to traffic control. The general concept and challenges of self-organization are reviewed, as well as analytical approaches involved in self-organized traffic control systems. Section V looks at the driving forces for these principles: new capabilities in data collection and communication. Section VI examines design elements of self-organizing systems. Section VII provides concluding remarks.

## II. TRAFFIC AND TRAFFIC CONTROL MODELS

Traffic control systems use a variety of strategies to determine cycle lengths[1], phase splits[2] and offsets[3]. One commonly used traffic control system is the Urban Traffic Control System (UTCS) developed by the U.S. Federal Highway Administration (FHA) in the 1970's. In UTCS, timing is optimized[4] for each signal based on average traffic conditions at a given intersection. Signal timing schedules are then developed and periodically downloaded to each controller. The objective is to maximize the bandwidth of major arteries and minimize delays. Given the unpredictability of traffic patterns, however, this technique is obviously suboptimal. In general, constant variations in the number and speed of vehicles make it difficult to account for all possible combinations during optimization exercises.

*2.1 Prevalent Adaptive Traffic Control Systems*

Real-time adaptive traffic control systems that incrementally compute local conditions were developed to address inefficiencies in the UTCS. The most notable adaptive systems are: Optimized Policies for Adaptive Control (OPAC) [32]; [38]; the Split Cycle Offset Optimization Technique (SCOOT) [39]; [40]; [41]; the Sydney Coordinated Adaptive Traffic System (SCATS) [42]; [40]; [43]; and the Real-time Hierarchical, Optimized, Distributed, Effective System (RHODES) [44]; [45].

OPAC was first developed at the University of Lowell in the early 1980s, and has evolved over the years [46]. It is a real-time distributed signal timing system based on measured and predicted traffic demand. OPAC does not use the concept of cycle; instead it optimizes control periods for particular intersections through phase switches at fixed time intervals. At each switching sequence a delay function is defined based on the initial queue length times the interval of time in which the cars are queued, with the goal of minimizing total vehicular delay [38]. For each stage (a sequence of one and no more than three phases) the optimal switching decisions are calculated independently using on-line data from upstream detectors as well as historical data. The optimization procedure is essentially a constrained search method [46].

SCOOT [39], developed in the UK, uses a steady-state model of traffic flow and platoon-dispersion equations to minimize queue lengths and delays in the grid. SCOOT gathers data about approaching vehicle platoons from upstream detectors in order to update the offset and cycle time at intersections downstream. Data is continuously collected and stored in a processed format by ASTRID, the "Automatic SCOOT Traffic Information Database." Relevant data include: the number of vehicles per hour arriving at the intersection: total delay in vehicle hours (average queue length in vehicles on a link); congestion; detector flow (recorded as vehicles cross a detector); and detector occupancy (the total number of quarter-seconds that the detector is occupied as a percentage of the whole period). Data is collected at a number of levels, including detector, link, node, region, route, and area.

---

[1] Cycle length is total time required for the entire phase set at an intersection where a phase set is the set of unique combination of vehicle or pedestrian phases that occur at the intersection
[2] Phase splits are the portion of the cycle length devoted to different phases during a complete cycle
[3] Offsets are the time delay in start of signal cycle across two adjacent intersections typically used to coordinate traffic flow across multiple signals.
[4] We use the optimized as a measure of improvement of the fitness function. Given the complexity of the problem a global optimum is not guaranteed or even achievable.

Real-time data collected by SCOOT is used to generate traffic flow models, or "cyclic flow profiles." These profiles are used to estimate the number of vehicles arriving at intersections downstream, providing estimation of queue size for hypothetical changes in signal timing parameters. Prior to each phase change the flow model is used to determine whether to delay or advance the time of the phase change by 4 seconds, or leave it unaltered. Similarly, once in each cycle, a decision is made to either advance or delay the offset, or to leave it unaltered. Finally, once every few minutes, a decision is made to determine if cycle time should be increased or decreased.

SCATS [47], developed in Australia, is a dynamic control system with a decentralized architecture that updates intersection cycle length using detectors at a stopline. In this system, offsets between adjacent intersections are predetermined and adjusted based on sensor information collected in real-time. The SCATS model groups intersections into subsystems, each subsystem containing only one critical intersection. The critical intersection coordinates all other intersections in the subsystem. Each intersection, at the subsystem level, can adjust the signal phase independently based on traffic conditions. Moreover, time saved from any phase change is included in subsequent phases in order to maintain a common cycle time for all intersections in the subsystem.

SCATS uses as its basic data the *degree of saturation*, or the ratio of the actually used green time to the total available green time. The cycle time of a critical intersection is periodically adapted to preserve a high degree of saturation, and the system also strives to maintain an equal degree of saturation on competing approaches by adjusting phase splits at the critical intersection. Although degree of saturation is used to improve phase timing, the SCATS system does not explicitly attempt to optimize any performance measure such as average delay or stops.

RHODES [44] is a traffic control system that predicts traffic flow in real-time, and optimizes phase timing based on a chosen performance measure, such as average delays, stops and throughput. Its decision system employs a hierarchical control structure. The top layer includes a *dynamic network loading* model that optimizes traffic based on slowly changing characteristics. It calculates aggregated flow data of the entire network, enabling it to make initial estimates about the load on each road segment. The middle layer, called *network flow control,* optimizes the flow across individual arteries, altering the phase and cycle time to accommodate the movements of platoons. The lower layer optimizes the flow based on individual vehicles at intersections.

## 2.2 Additional Traffic Control Models

A number of other models for traffic control systems have been proposed in the literature, employing a variety of techniques. Dynamic programming and decomposition techniques were used with DYPIC [48] and in the 1980s with the PRODYN algorithm [49]. PRODYN uses dynamic programming in a hierarchical fashion to make signal decisions at the intersection level, while using a decomposition coordination technique at the upper level of the traffic network [49]. Early work in field testing adaptive algorithms is exemplified by Vincent and Young's Microprocessor Optimized Vehicle Actuation (MOVA) strategy, which works by measuring the benefits of extending a green phase [50].

Mauro and Di Taranto [51] described a hierarchical system that applies priority assignment to public transport and achieves global optimization. In their approach, called Urban Traffic Optimization by Integrated Automation (UTOPIA), the decomposition algorithm considers an "intersection" and an "area" level, each with two main parts: an observer and controller. At the intersection level, the observer estimates intersection state based on traffic counts and state of traffic light; the controller determines the signal settings. At the area level, the observer obtains actual traffic counts from the network based on a time discrete model. The controller optimizes the network through the *rolling horizon* technique, and sends information to the intersection level. The system also gives absolute priority to selected vehicles. In field experiments, the model showed good results for a number of criteria.

Yagar and Han [52] also proposed a real-time algorithm that takes into account the presence of transit vehicles (*i.e.* vehicles that frequently need to load/unload passengers) in order to better allocate priority levels. In their approach, detectors upstream capture information about vehicle type as well as time of detection, and create a priority list. Based on this information, the algorithm attempts to minimize cost of stops and delays. Diakaki and colleagues [53] present an extension to the Traffic-responsive Urban Control (TUC) model, a signal control strategy well-suited for saturated traffic conditions. In this extension, the TUC is able to perform real-time cycle and offset control, also enabling priority consideration for public transport.

ALLONS-D, presented by Porche and LaFortune [54], uses a decentralized architecture to provide real-time adaptive signal plans at the level of the intersection. It uses a tree searching algorithm to minimize aggregate delay at all approaches to an intersection [54]. Unlike the SCOOT and SCATS systems discussed above, ALONS-D is not constrained to create a cyclical plan. This non-cyclic approach is also a feature of UTOPIA, OPAC and PRODYN, all previously discussed. ALLONS-D uses a rolling horizon approach to determine vehicle arrivals, but unlike OPAC and UTOPIA, for example, the horizon is defined by current traffic conditions and the time to clear all vehicles on the intersection as identified by upstream detectors [54].

Gradinescu and colleagues [55] propose a system that can make signal control decisions that localize vehicles in the traffic system through wireless devices that may be present inside the vehicles. The control system is supported by a mobile ad-hoc network that is generated through the short range wireless signals around an intersection. With access to this information, the controller can measure volume and demand of traffic and determine cycle length through Webster's formula based on the critical flow per capacity ratio.

Tian and Urbanik [56] developed a signal timing approach that optimizes bandwidth efficiency. The approach adopts a partitioning technique that subdivides the arterial system into subsystems, and applies a bandwidth optimization algorithm. Signal timings are optimized at the subsystem level, and then combined to propose a system progression band for one direction, such as peak-traffic direction. The approach is able to maximize bandwidth efficiency in one direction, but such maximization is not guaranteed in all directions or subsystems.

A number of other distinct approaches to optimization may be found in the literature. Bazzan [57], for example, uses agent based modeling to propose a decentralized system that is able to coordinate and cooperate by using techniques from evolutionary game theory. Zhu and colleagues [58] use reinforcement learning to optimize network traffic. Sanchez, Galan and Rubio [34] [59] use a genetic algorithm to optimize the length of each cycle stage (the set of states of traffic lights in an intersection) measured by the number of vehicles that ultimately leave the network. Sanchez and colleagues [34] [59] also provide a modified version of a Cellular Automata traffic model that allows for the identification of various vehicle properties and rules such as position, path, smooth braking, and others. There have been numerous attempts to make real-time traffic controllers using techniques such as fuzzy logic, genetic algorithms, and neural networks. These techniques will be further discussed in section 3.

## 2.3 Urban Traffic Models

Given the complexity of urban traffic, urban traffic simulations are required to test traffic signal control methods. Several models for studying traffic behavior have their origins in physics, including: fluid dynamic models [60]; [61]; vehicle-following models [62]; [63]; and coupled lattice models [64]. A cellular automaton for a self-organized traffic grid in which signals communicate with immediate neighbors to coordinate actions is employed by both Biham et al. [65] and Ohira [66].

The majority of traffic models are intended for studying highways. Some highway traffic models have been extended to include intersections, while others have been proposed natively with intersections. Urban traffic models can be categorized by their level of detail as *macro-*, *micro-* or *mesoscopic* [67]. Macroscopic models consider traffic as a general, continuous fluid, such as the one presented by Lighthill and Whitham [60]. Microscopic models consider individual traffic entities (*e.g.* vehicles) as discrete units, and with particular spatiotemporal behavior. Mesoscopic models consider the behavior of individual entities but do not distinguish and trace their particular actions. As Hoogendoom and Bovy [67] indicate, there are other criteria by which to classify traffic models, including the dimension captured (*e.g.* velocity), scale (*e.g.* continuous, discrete), process representation (*i.e.* deterministic or stochastic), operationalization (*i.e.* analytical or simulation) and area of application, which refer to different types of an urban traffic network that may be examined (*e.g.* cross-section, single lane stretch). In this section, we review a selected number of models that illustrate some of the different approaches taken in the literature.

Macroscopic models were the first urban traffic models to be developed, as more detailed models require digital computers to handle their large number of variables. These models captured aggregate driver behavior and conceptualized them as continuous flows, such as fluids or gases. Examples of these hydrodynamic approaches include the Lighthill and Whitham [60] model and the Payne model [68], as well as the models of Philips [69], Kuhne [70] and Kerner et al. [71]. Examples of mesoscopic models include the *headway distribution models*, which use measures of the difference in time between two successively passing vehicles, such as Branston's generalized queuing model [72]; *models*, which identify groups of vehicles that share some properties, often using size and velocity as principal characteristics [73]; and *gas kinetic models*, which use velocity distribution functions to model traffic flows (e.g. see [74]). Hoogendoorn [75] unifies a number of gas-kinetic approaches to propose a generic model that describes the dynamics of vehicles based on the distinction between free-flowing versus platooning.

A number of microscopic models have been proposed in the literature (for a more extensive review, see [67]). Here we review a few microscopic models based on cellular automata. Cellular automata approaches have been especially useful for self-organizing systems, and their relatively simple principles offer an opportunity to model large, decentralized systems of cells, where the cells interact with their neighbors and adapt their behavior.

Nagel and Schreckenberg [76] present a cellular automata model of highway traffic flow (NaSch) in which a lane is represented by a one-dimensional lattice (i.e. a row of connected points). Each of the lattice sites represents a cell that can be either empty or occupied by at most one vehicle at a given time. This is a complex self-organizing system in which competing forces are at work keeping the system dynamic, but also organized. The rules incorporate acceleration, deceleration, randomization, as well as the actual movement of the vehicle.

In the NaSch model, vehicle motion is not deterministic; there is randomness due to unique driver reactions to traffic conditions, which adds a layer of complexity to the dynamics. The model suggests that, when both critical density and randomness are reached, spontaneous traffic jams will form. An interesting feature of this model is the flux density curve. The flux, or flow rate, is simply the number of vehicles per second at a particular point. At low densities, vehicles are free from interaction and can move at any speed. At high densities, vehicles must follow a strict order, as there are velocity oscillations due to reduced space, in some cases with vehicles having to stop. At medium densities, the interaction becomes most varied. Vehicles are close enough to interfere with one another, but not close enough to induce a high degree of order.

Biham, Middleton and Levine [65] present a model (BML) based on CA and a two-dimensional plane. In this model, a square lattice represents East-West and North-South streets. The spaces on the lattice represent intersections. The rule in this model is very simple: a vehicle moves forward by one lattice if and only if the site in front is empty. At a non-zero density, a phase transition can occur in which the speed of the vehicles discontinuously approaches zero. This is due to a traffic jam so complete that it blocks all vehicles in the system. The BML model assumes random initial distribution of vehicles, but subsequently provides a fully deterministic representation.

It is possible to combine the NaSch model, which describes traffic flow, with the BML model, which includes traffic control representation at intersections. A transition from the freely flowing phase to the completely jammed phase has been observed in such models at a vehicle density that depends on $D$ (the number of points in the linear lattice for the NaSch model) and $T$ (the traffic light cycle length). The models are clearly simplifications, and purposely leave out such details as the number of lanes, traffic light configurations, dedicated turn lanes, and left and right signals. Nagel and Schreckenberg's demonstration that, at certain densities, traffic patterns change from laminar flow to start-stop waves has also been used to improve traffic control systems. Krauss, Wagner and Gawron [77] (1997) present a generalization of the NaSch model characterized by a family of models with one parameter $r$ (the time steps that cars take to come to a complete stop from maximum velocity). Their model properly accounts for the discontinuity and the different densities that occur when traffic breaks and a jam occurs [77].

Chowdhury and colleagues [78] have examined the implications of both the NaSch and BML models, and note that one cannot assume that traffic is uniformly distributed; if it were, traffic light optimization would be the same for all lights. Instead, some roads dominate others in terms of traffic load, which influences local conditions.

Rickert, Nagel, Schreckenberg and Latour [79] provided an extension of the initial NaSch model to include multi-lane traffic. Fouladvand and colleagues [80] modelled traffic flow on roundabouts. Here, traffic is controlled by a self-organized scheme in which lights are absent. This controlling method incorporates a yield-at-entry strategy for the approaching vehicles to the circulating traffic flow in the roundabout, also using the stochastic cellular automata approach of the basic NaSch model.

An interesting experiment by Brockfeld and colleagues [81], examining traffic at a single intersection, reveals that throughput is a function of cycle length and reveals an oscillatory pattern. Their research also shows that experimenting with green wave synchronization of multiple lights is not likely to achieve any additional increase in throughput. Additional experiments with green wave synchronization were conducted by allowing a phase offset between light cycles, allowing minimal delay of traffic flow due to red lights. The outcome, as expected, is very similar to the previous result for a single intersection in terms of flow rate versus traffic light cycle length. These experiments also indicate that random traffic light cycle offsets, for low cycle times, outperform green wave traffic flow.

Burghout and Wahlstedt [82] investigate a hybrid traffic simulation model that integrates VISSIM (a microscopic model) and MEZZO (a mesoscopic model). The hybrid model is applied to a network in which MEZZO simulates the entire Stockholm area (6000 links), and VISSIM simulates a small area of interest containing three intersections with adaptive signal control and bus-priority functions. The control and functions are simulated by a separate signal controller that would take place in the field. They evaluate a fixed-time and adaptive control and show clear improvement in terms of travel times, delays and stops. They see additional traffic in the local area but an overall reduction in congestion, demonstrating the merit in simulating local and global traffic grid simultaneously. A self-organized system will similarly model local control and evaluate global effect on the entire grid. Winters et al. [83] examine the Python-based Lightweight Intelligent Traffic Simulator (LITS). The LITS is a vehicular network simulator that models vehicle to traffic light communication with a goal of providing a test engine for other researchers to simulate their algorithms.

Considering the number of factors that should be taken into account when modelling urban intersections, and the difficulty in properly calibrating and validating models, Pretorious et al. [8] recommend evaluating intersections simply in terms of volume/capacity ratios, or the saturation level, and carrying out improvements on the basis of minimizing this ratio. Similarly, Akgungor and Bullen [84] suggest that saturation measures are the most important to assess delay. Delay in turn serves as the most important measure to assess quality of service at an intersection. Akgungor and Bullen [84] propose improvements in traffic modeling by presenting a methodology for determining the delay parameter *k* as a function of the degree of saturation, rather than using a fixed value. In related work, Balmer, Axhausen and Nagel [85] present a flexible agent-based modeling approach that allows for a variety of input data to create large-scale transportation simulations. Suzuki et al. [224] introduced an Ising-like traffic model, where the state of traffic signals interacts locally and leads to chaotic dynamics.

Rosenblueth and Gershenson [86] provide a CA model that is simple yet realistic, as it takes into consideration the effect of traffic lights. Many of the basic approaches to urban traffic modelling, including the NaSch and BML models, have seen updates to include additional realistic conditions of traffic (*e.g.* see [78]). Moreover, it should be noted that the performance of the *testing tools and platforms* may vary depending on the topography, traffic density, data available, and user expertise. They also vary in the interfaces used and the optimization techniques employed, but they all tend to use standard traffic models available in existing literature.

## III. REVIEW OF ANALYTIC TECHNIQUES USED IN TRAFFIC CONTROL OPTIMIZATION

This section reviews the large body of work on traffic signal synchronization. It focuses specifically on the analytic techniques branch of the Traffic Signal Control Taxonomy presented in Figure 1, namely: dynamic programming, control theory, fuzzy logic, machine learning, and queuing theory. Each of these elements will be discussed in turn in this section.

### *3.1 Dynamic Programming*

*Dynamic programming* was introduced by Bellman [87] as a way to solve multistage decision problems. It is a way to reduce the problem into simpler sub-problems, and then combining the solutions to get the solution to the original problem; the approach examines all possible ways to solve a problem and then selects the best solution. This is a time-consuming and slow approach, but one that guarantees an optimum solution. Often, algorithms that do not require a brute-force approach like a greedy algorithm are preferred to reduce the computational burden.

Dynamic programming has been applied at least since the 1970s with DYPIC [48], and the 1980s with the PRODYN algorithm [49]. More recently, Heung et al. [27] introduced dynamic programming for signal optimization at intersections with a fuzzy-logic controller. The assignment of green time to each phase of a traffic cycle is considered a multistage control problem with a finite number of possible actions at each stage. A dynamic-programming technique is used to derive the green time for each phase in a traffic cycle. Projected traffic flow from adjacent intersections is incorporated into the analysis to ensure that control is extended beyond a single junction. The controller does not require complex hardware, and the simulation results show that the delay per vehicle can be substantially reduced, even when traffic demand reaches capacity. This bodes well for a self-organization approach, where information from adjacent intersections is considered during the optimization process.

Yi et al. [88] use dynamic programming to optimize flow at intersections during periods of oversaturation. They also address the problem of detection in oversaturated conditions when the queue spills past the detector. The authors add a detector that counts vehicles leaving the intersection, allowing for a better estimate of queue size. Dynamic programming leverages the fact that if a state-action is optimal, then removing the first state and action still leaves an optimal state-action sequence [89]. This is known as the Bellman Optimality Principle.

In other research, Cai [90] and Cai, Wong and Heydecker [91] build on the work of Allsop [92], and provide an *Approximate Dynamic Programming* (ADP) framework in which control parameters can slowly change over time through learning techniques. ADP addresses the computational burden required from dynamic programming, as well as the need for complete information by simplifying the cost-to-go function. A machine learning technique, such as reinforcement or perturbation learning, can be employed to adjust the functional parameters.

*3.2 Control Theory*

Traffic signal optimization can be framed as a control problem. However, application of *control theory* can vary based on the input parameters and the traffic parameters being controlled. Classical control theory involves a closed-loop system with sensors and feedback. The feedback is used to control states or outputs; input parameters (*e.g.*, the number of cars waiting at a red light) have a direct effect on control parameters (*e.g.*, adjusting traffic light cycle time). The input parameters are measured with sensors, and are processed by the controller. The result (the control signal) is used as input to the process, thereby closing the loop. In an open-loop system, there is no sensing or feedback from output, which makes it less robust than a closed-loop system. An example of an open-loop traffic light system is one in which historical traffic patterns are used to set the timing parameters; as long as the traffic flow follows the historical trend, the system performs well. However, variations from the norm, which can be expected, can lead to significant performance deterioration. A closed-loop control system will typically use both historical traffic data and real-time traffic conditions to make decisions.

Shimizu and colleagues [93] frame traffic signaling as a classic feedback control system to minimize congestion at intersections. The input is the number of cars approaching the intersection, and the control variables are cycle length, green split, and cycle offset. Oversaturation and queue spillback both adversely impact the performance of the control algorithm. Wey [94] uses the Proportional Integral Derivative (PID) for signal control to minimize traffic delay. Ball et al. [95] have developed the H∞ method to determine optimal signaling control for an intersection. The reference point keeps the queue length below a predefined maximum value. This technique assumes some means of observing the vehicle flow rate. The effective green time and total cycle length are the control variables. The basic of the H∞ features are presented in an article by Kwakemaak [96]. Feedback control may also be used to order traffic flow. Yang and colleagues [97] apply linear feedback control to signal setting, using reserve capacity and degree of saturation as the control objectives.

Another nonlinear control optimization technique called *adaptive fine-tuning* addresses the problem of efficiently determining an optimal value by evaluating the objective function on random perturbations around the current control parameter vector [98]. The results are similar to gradient descent. This fine tuning algorithm has been applied to a traffic signal control strategy called traffic-responsive urban control (TUC), which allows for control of green times, cycle times and offsets; it also allows for the provision of public transport priority, and uses the store-and-forward concept described below.

In the work of Aboudolas and colleagues [99], a linear-quadratic optimal control is applied to the store and forward model (SFM)—not to be confused with the notion of store and forward in computer networks. Aboudalas and colleagues' model achieves a significant reduction in complexity by making the discrete time step equal to the traffic light cycle time. This means that the model only samples the average output from a full traffic cycle, and ignores the intra-cycle timing and queue variations, thus simplifying the model at the cost of some high-resolution variation. Pecherkova and colleagues [100] meanwhile, focus on state and parameter estimation techniques for a traffic signal control system, examining two techniques: the unscented Kalman filter and divided difference filters.

## 3.3 Fuzzy Logic

The concept of *fuzzy logic* was first introduced by Zadeh [101] in "Fuzzy Sets." Fuzzy logic allows for uncertainty in representation of information by allowing partial membership of sets. It mimics human perceptions in control logic ,making it effective in representing human behavior. Pappis and Mamdani [102] applied fuzzy logic to the analysis of traffic control. Their model assumes that vehicle loop detectors are placed upstream of the intersection to measure approach flows and estimate queues. This information is used to determine whether to extend or terminate the current signal phase. Each possible extension is assigned a degree of confidence, and the extension with maximum confidence is implemented. Before the implementation ends, the rules are reapplied to see if further extensions are required.

The goal of fuzzy logic is to approximate expert perceptions. A fuzzy controller consists of a system that contains fuzzy rules, a database that collects traffic data, and an inference engine that analyzes the rules. Initially, fuzzy systems used a static set of rules, but more recent versions use evolutionary principles (*e.g.*, genetic algorithms) to refine the set of rules governing traffic flow [103]; [104]; [22]; [105]; [106].

Modern fuzzy systems assess the fitness of the initial rule set and select the most appropriate for propagation into the next generation. The genetic operators of crossover and mutation are applied to those selected, and a new generation of rules is obtained. With each generation, the average fitness of rules is expected to improve, resulting in better overall performance. Niittymacki and Kikuchi [107] use fuzzy logic to control a pedestrian crossing signal, while a number of other researchers have investigated decentralized approaches that incorporate self-organization concepts and fuzzy rules [23]; [108]; [109]; [110].

*3.4 Machine Learning*

The simulation and optimization of traffic light controllers using an adaptive optimization algorithm based on *reinforcement learning* is analyzed by Wiering and colleagues [111], as well as Cai et al. [91] as explored earlier. Wiering and colleagues examine a traffic light simulator that enables experimentation with different infrastructures and compares different controllers. Their experimental results indicate that the adaptive traffic light controllers outperform other fixed controllers in all infrastructures.

Neural networks—a type of machine learning technique—has been used to improve the quality of fuzzy rules for traffic signal optimization in a number of studies [112]; [113]; [114]. Researchers train the network to adapt to different traffic flow conditions based on real-time data. Elements of dynamic programming have also been added to form a neuro-fuzzy controller for improving the performance of the traffic optimization model across multiple networks [115].

A hierarchical architecture that combines neural networks, fuzzy logic, and genetic algorithms to create a multi-agent architecture system has been constructed by Choy and colleagues [116] [117]. The authors divide the primary problem into sub-problems, each of which is managed by an intelligent agent using fuzzy neural decision-making rules. A reinforcement learning process is implemented to improve the quality of decision making. Zhu et al., [118] use a radial basis function neural network to forecast traffic volumes at intersections. The unique contribution that they have is that they provide information about adjacent signals (15 min intervals) to the neural network for improved accuracy. The question remains how this approach will work across different settings, especially those where volumes vary significantly across multiple intersections.

Lu [119] has demonstrated the applicability of incremental multistep Q-Learning to adaptive traffic signal control. Q-learning is unsupervised learning, and based on trial and error. Rather than being presented with a large set of training examples, the Q-learning agent essentially generates its own training experiences from its environment. Ozan et al. [120] have used Q-learning in conjunction with a TRANSYT traffic simulator and shown improvement compared to Hill Climbing and Genetic Algorithms. Their benchmark using standard test cases demonstrates the feasibility of using reinforcement learning in real-time traffic applications. Zhu et al. [58] develop the Junction Tree Algorithm (probabilistic graph-based algorithm), which is another reinforcement learning algorithm that outperforms independent learning (Q-Learning), real-time adaptive learning, and fixed time algorithms across multiple performance metrics. Their simulation was done on 18 signalized algorithms in VISSIM.

*3.5 Queuing Theory*

Pioneering work in *queuing theory* started with Webster in 1956, and has since evolved into a widely accepted analytical tool for traffic modeling. Queuing theory applies to the queue of cars that form at the intersections. The analysis of traffic flow for signaling has taken two general approaches: a *deterministic* fluid flow component assuming continuous variables and steady flow; and a *stochastic* component comprised of probabilistic events and their impact on queuing. Queuing theory works well for steady state, but becomes problematic when the traffic flow rate exceeds capacity. Another complicating factor is the lack of independence of traffic lights. While standard queuing input rate assumptions can be assumed at an isolated signal, systems of interconnected signals impact one another in ways that violate the Poisson assumptions [121].

Queuing formulae for traffic control can generally be separated into an under-saturated, or uniform, queuing component and an oversaturated, or stochastic, queuing component [9]; [122]. The uniform component assumes that traffic arrives and departs in a uniformly distributed manner. Yet, since real traffic is not uniform, a stochastic component must be included in the analysis. When the load approaches capacity, the stochastic effects tend to dominate.

During transitional periods, the standard deviation can be of equal magnitude than the mean.

Beckmann and colleagues [123] provided the first queuing derivation for traffic signal delay under the assumption of binomially distributed arrivals and uniform service rates [123]. Webster [124] applied a different formula comprised of three components: (1) an analytical derivation of uniform delay; (2) a characterization of stochastic delay analytically derived assuming Poisson arrivals and deterministic service rate; and (3) an additional term used to match the analytical results with those observed from simulation data. Webster's delay is still widely used and continues to serve as the benchmark for traffic signal control.

Micro-simulation models are normally used to evaluate traffic-adaptive signal control systems. Lehmann [125] developed an analytical approach to this evaluation based on queuing models. In particular, he investigates a queuing model for a simplified adaptive control strategy based on a rolling horizon scheme, in which a signal serves two movements alternatively. A numerical algorithm is developed in a stochastic context to compute steady-state performance measures such as average delays and expected queue lengths. These results are compared with simulation-based results; the analytically derived numerical method predicts the simulation results quite well. Foulandvand, et al. [225] proposed an early decentralized approach for regulating traffic lights based on queues. However, their local optimization failed to improve traffic globally. Still, a similar approach was later used by self-organizing methods (detailed in Section IV), which successfully improved global traffic.

Tian [126] proposes two models that estimate the capacity of an intersection using actuated control: the Minimum Delay Model and the Hybrid Model. In these models, the capacity of an approach to a lane group of the intersection is a function of the saturation flow rate, the green time allocated to this approach or lane group, and the cycle length of the intersection.

The Minimum Delay Model estimates the green times and cycle lengths from flow rates, minimizing total delay. Parameters in this model include the ratio of green extension period to queue service time specific to each approach or lane group. The parameters depend on the distribution of arrivals at the intersection. The Hybrid Model combines the deterministic queuing model that estimates queue service time, and a theoretical model that estimates the green extension period from the unit extension, the flow rate, the speed limit of the approach, and the detector length. A method converting the left turn traffic volume to equivalent through volume is developed. The method is used to estimate the capacity of intersections which permit left turn phases.

The Minimum Delay Model and the Hybrid Model are validated at the intersection level by comparing the estimations of effective green ratios with those simulated by MITSIMLab. These two models have been validated at the network level with real data. The results show that both the Minimum Delay Model and the Hybrid Model are appropriate for estimating capacity of intersections with actuated control. The Minimum Delay Model is also suitable for estimating capacity of intersections with adaptive control. The Emulation Model is applicable to off-line mesoscopic dynamic traffic assignment.

A technique loosely related to queuing theory, known as *shockwave analysis*, uses the notion of vehicle arrival rate, but differs in the notion of service rate. Instead of the paradigm of a server and service rate with an associated queue for vehicles waiting for service, vehicle density and flow rate changes are monitored. The ratio of change in flow rate to change in density yields a wave velocity that must have occurred to account for the changes in flow rate and density.

Shockwave analysis uses an intuitive, physics-based derivation, while queuing theory utilized a more computer communication network or operations research-based derivation. Yeon and Ko [127] discuss the relationship between shockwave analysis and queuing theory, and attempt to experimentally validate the comparison using a real traffic flow study. Neither approach appears to be consistently more accurate than the other in the study. Le, et al. [226] propose a decentralized BackPressure scheme, where queue length measurements "press" on downstream queues. In this way, streets with a higher demand will obtain higher allocated green times within fixed cycles.

Freeman et al. [128] benchmark different tools available for traffic optimization, including Highway Capacity Software, PASSER, SYNCHRO, SimTraffic, TRANSYT, and CORSIM. Their work focuses on accurately predicting delays, queue lengths, spillback, and levels of service. Sensitivity analyses are performed to evaluate limitations. Their aim is to identify the requirements for an "acceptable technique" to aid in the design and evaluation of signalized intersection systems.

The fundamental limitation in previous work is both the choice of techniques that are used, and the formulation of the problem. Formulating the traffic signal optimization as a centralized control problem does not lend itself to a solution that is either robust or scalable. As the number of variables grows, the complexity of the problem increases, making it infeasible to solve. Also, the techniques that rely on static data based on historical traffic patterns are unable to capture the instantaneous variability of the traffic flow, making the solution suboptimal. In addition, the techniques are not robust to traffic anomalies, such as sudden influx of traffic or the disruption of an artery. As an alternative to overcome these limitations, we examine the use of self-organization as a viable option for traffic light coordination.

# IV. SELF-ORGANIZING TRAFFIC SIGNALS: AN ADAPTIVE AND EFFICIENT APPROACH

There are several instances in nature in which groups of organisms work together to collectively accomplish complex tasks that they would otherwise be incapable of performing individually. In these instances, without the need of central control, each organism has the capacity to coordinate with its neighbors, leading the group to function seamlessly in a well-organized fashion. Consider, for instance, the ant. It is a "simple" biological organism, in the sense that its behavioral repertoire is limited to 10-40 elementary behaviors. In groups, however, ants exhibit sophisticated collective behavior including the ability to build rafts and bridges by tangling their legs together. None of the insects grasps the big picture, but collectively they can perform sophisticated tasks that contribute to the success of the overall colony. This strong integration has led some biologists to consider the colony as an organism in itself.

The secret lies in self-organization: individuals follow simple rules based on local information, collectively completing complex tasks. Such behavior is also called *swarm intelligence* [129]. This behavior is not only observed in social organisms, but also in physical phenomena such as the crystallization of liquids [130] and the alignment of magnetic spins [131]. Self-organizing techniques leverage the interaction of many similar components. Each component follows a set of rules and adapts its behavior by sensing its neighbors and/or local conditions. Simple behavior at a local level leads to complex collective outcomes; the neighbor-to-neighbor interaction leads to global coordination of the entire group. Self-organization may thus be defined as an increase in order without outside control.

A rich literature has developed around the general concept of self-organization [132]; [133]; [134]; [135]; [136]; [37]; [137]. In studies of self-organized systems, entropy is used as the most common definition of disorder. The usual approach is to look for conditions that appear to violate the second law of thermodynamics, which states that systems in closed systems evolve to reach thermal equilibrium, or maximum disorder. Such a violation was posed by Maxwell's demon in relation to adiabatic computing. Self-organizing systems do not violate the second law of thermodynamics, and thus must constantly ingest matter or energy with low entropy and expel the high entropy as heat. The information-theoretic view of the second law may be stated as the fact that every system tends to its most probable state [138]; [139].

## 4.1 Philosophical and Practical Challenges

Self-organization in multi-agent systems appears to contradict the second law of thermodynamics. This paradox has been explained in terms of a coupling between the macro level that hosts self-organization (and an apparent reduction in entropy) and the micro level (where random processes greatly increase entropy). Metaphorically, the micro level serves as an entropy "sink," permitting overall system entropy to increase while sequestering this increase from the interactions where self-organization is desired. Wiering and colleagues [111] make this metaphor more precise by constructing a simple example of pheromone-based coordination, defining a way to measure the Shannon entropy at the macro (agent) and micro (pheromone) levels, exhibiting an entropy-based view of coordination

There are many philosophical open questions related to self-organization that restrict its use in engineering. For example, it is the observer who ascribes properties, aspects, states, probabilities and, therefore, orders to the system. But organization is more than just low entropy: it is structure that has a function or purpose. Thus, Stafford Beer (1966) made a very important observation when he noted that what under some circumstances can be seen as organization, under others can be seen as disorder, depending on the purpose of the system [138].

One of the most common problems associated with self-organizing systems is emergence. Self-organizing systems typically have higher-level properties that cannot be observed at the level of the elements, and that can be seen as a product of their interactions (more than the sum of the parts). Emergent properties are a hallmark of complex systems. The problem here is ontological. According to classical thought, there is only one "true" description of reality. In this case, a system cannot be a set of elements and a whole with emergent properties at the same time. Still, there are alternatives to classical thought that allow an understanding of emergence [140].

One solution is to distinguish between an absolute "god-like" being that observes the system, and mere mortals who possess only limited perception. While this kind of discussion is philosophically interesting, it is of limited use in engineering. At some point, one must choose a practical definition and stick with it. The one that we have chosen is expressed by Gershenson as follows: "A system described as self-organizing is one in which elements interact in order to achieve dynamically a global function or behavior." [36]

Traffic grids have the basic components needed in a self-organizing system: traffic lights are distributed through the grid; there is connectivity between intersections via streets; and there can be local communication (at least in terms of cars flowing from one intersection to the next). Consequently, traffic optimization can benefit from self-organization models such that traffic signals communicate with their neighbors and adapt their signal timing accordingly. Traffic signals adapt their behavior based on a universal set of rules using local information, and all the intersections can be coordinated as changes cascade through the entire grid.

*4.2 Techniques for Self-Organized Traffic Control Systems*

There are several techniques on the basis of which self-organized traffic light control systems can be created. Some of the most effective self-organization models are built using a set of heuristic rules developed from observation of natural phenomena, that are then abstracted and adapted to the context of a specific problem (*e.g.* [141]). The goal is to come up with a set of rules to govern the behavior of individual components of a self-organizing system that will result in an overall coordination of the entire system. Ideally, we would like to invert the problem and derive the set of rules based on the desired outcome [142]. However, there is no closed-form solution, and the heuristic rules are created based on an initial intuition and then refined based on observation using trial-and-error.

There have been many formal ways to approach the problem. One formal technique that has been used to investigate complex interaction networks is *graph spectral analysis*, which relates the network topology of the road system to fundamental mathematical properties such as connectivity and maximum flow. Eigenvalues of the adjacency matrix of the road network produced in this type of analysis have many important properties that have not yet been used in traffic signal synchronization. Pulse-coupled oscillation (PCO) [143] and Markov Random Fields (MRF) [144] are two other self-organizing techniques. PCO is the mechanism behind firefly synchronization; fireflies adjust the timing of their flashes to the energy pulses of their nearest neighbors. Our own research has examined the possibility of using PCO to provide synchronized, low-power time pulses for sensor networks. Similar techniques may be used to achieve new types of green signal waves, as proposed by Faieta and Huberman [145]. The use of MRF, meanwhile, achieves global coordination by minimizing local energy functions. This technique comes from the Ising model used in physics for applications such as image processing. Spatial dependencies can be utilized to fill in missing or corrupt images. MRFs are also used in forward error correction in communications. Similar techniques might be used to set traffic light timing based upon spatial configurations.

Self-organized traffic-flow models have been investigated using cellular automata [76], swarm algorithms [146], coupled phase oscillation [147], statistical physics [148], [149], fluid dynamics [150], queuing theory [151], multi-agent simulations [36], and fuzzy logic [23]; [24]. Similar complex systems techniques have also been applied to modeling pedestrian flows [152] and traffic jams [153]. We discuss these and other works in greater detail in the following subsections.

*4.2.1 Heuristics-Based Algorithms*

Current *advanced traffic management systems* (ATMS) use learning methods to adapt phases of traffic signals,

normally using a central computer. This self-organizing approach is inherently distributed, as the global synchronization is adaptively achieved by local interactions between vehicles and traffic signals, generating flexible *green waves* on demand. The notion of green wave [154] is popular with traffic engineers: phases and periods of light signals are chosen such that a series of lights is coordinated to allow continuous traffic flow through multiple intersections, typically in one or two directions, as there are mathematical constraints to achieving full coordination in all directions. Green waves can be effective if traffic velocity is relatively constant, and most traffic goes in the same direction at the same hour. However, the optimal synchronization of traffic lights changes with vehicle density: as traffic density increases, vehicle speeds tend to decrease. In this way, different coordination schemes are required for different densities. To resolve these inefficiencies, numerous attempts have been made to synchronize traffic lights by using a central computer to collect and analyze data and adapt to real-time traffic conditions [39]; [155].

There are several problems with this approach, including latency and poor scalability. Current ATMS use learning methods to adapt traffic light phases with the help of a central computer. Still, this adaptation tends to be much slower (at the scale of minutes or hours) than the change in demand (at the scale of seconds). The self-organizing approach discussed here does not need a central computer, as global synchronization is adaptively achieved by local interactions between cars and traffic lights, generating flexible green waves on demand. Cools and colleagues [156] dispute the efficacy of green wave methods—including ones that have been tested in a number of simulations—such as the Green Light District/iAtracos project. Simulation results from Belgium indicate that a self-organizing controller performs better than the green wave controller. Further work has confirmed this conclusion [157]; [158]; [159].

Gershenson [36] and Gershenson and colleagues [156], have written extensively about self-organizing systems, specifically self-organizing traffic control systems. One aspect of optimal control that features prominently in their work is *waves*—a sequence of traffic lights in a single direction that allows for continuous flow through multiple intersections. Appropriate phases and periods that optimize traffic flow can be readily selected if a constant traffic velocity is assumed. However, as traffic density increases, speed tends to be reduced.

Gershenson also addresses self-organization and complexity from fundamental principles [36]; [160]. The algorithm he proposes is very simple and does not require direct communication, just the ability of each traffic signal to count the number of vehicles approaching the light while it is red [161]; when this count reaches a threshold value, the green signal cycles to yellow then red. The red signal resets the count to zero and proceeds as before. This tends to encourage larger groups of vehicles (convoys) to flow together. In a sense, this acts like congestion control in a communication network in which the maximum delay-bandwidth product of a link is utilized. There also tends to be larger spaces between the convoys, thus allowing for cross traffic, and there is a time threshold so that the light does not change too quickly during periods of high traffic density.

Two further conditions are taken into account to regulate the size of convoys. Before changing a red signal to green, the controller checks if a convoy is crossing through, in order to avoid breaking it. More precisely, a red light does not change to green if, on the crossing street, there is at least one vehicle approaching within a pre-specified distance of the intersection. This rule keeps crossing convoys together.

For high densities, this condition alone would cause havoc, since large convoys would block the traffic flow of intersecting streets. To avoid this, a second condition is introduced. Condition one is not taken into account if there are more than a specified number of vehicles approaching the green light. With this rule, long convoys can be broken up if a convoy will soon be through an intersection. The metric used in this approach is average trip waiting time, which is the travel time minus the minimum possible travel time averaged over all vehicles. Note that the algorithm parameters mentioned previously had to be determined manually. Vincent and Young [50] adopted a similar self-organizing technique decades before for isolated intersections. This algorithm has recently been refined with further rules which increase the performance for very low and very high densities, bringing them close to the theoretical optimum for all densities [158]. Burguillo, et al. [227] extended [160] to propose history-based self-organizing traffic lights. In this method, the duration of the next green cycle of every light is directly proportional to the number of cars which crossed the intersection in the last green cycle. Results showed slight improvements over [161].

Particle Swarm Optimization (PSO) is a meta-heuristic algorithm that does not require computation of gradients [162]. It changes the solution stochastically and uses an optimization function to evaluate the objective function. The algorithm has the capability of navigating local optima in the search space. However, there is no guarantee of a global optimum. Chen and Xu [33] apply PSO to traffic signal timing by deploying a local fuzzy-logic controller (FLC) installed at each intersection. These controllers provide some initial solutions for the algorithm; coordination parameters from adjacent intersections are incorporated, and the membership functions and rules of the FLC are optimized. They optimize the average delay and average number of stops for adjacent intersections. Simulation results show that the delay per vehicle can be substantially reduced under both constant and time-varying traffic demands, particularly when upstream demand exceeds downstream demand. The implementation of this method does not require complicated hardware; its simplicity makes it a useful tool for off-line studies and real-time control purposes. The rules, however, need to be made universal to work under different scenarios. Kammoun et al. [163] present an optimization algorithm for route guidance for vehicles based on GPS information of vehicles that incorporates fuzzy logic in decision analysis. In their formulation the road network is hierarchically organized with multi-level decision analysis.

Tatomir and Rothkrantz [164] propose a dynamic traffic routing algorithm based on Ant Colony Optimization (ACO) model. They use a routing table with potential split rates of traffic flow; at each network node the traffic is split into different arteries based on the split rates defeined in the table. Using a central table to define the splitrates results in poor scalability as the network size grows. Alves et al. [165] use ACO for traffic optimization for single-origin single-destination networks in which they use a static traffic model with time-invariant traffic conditions. Cong et al. [166] present their ant colony routing algorithm for redirecting traffic at intersections, and address the limitations that they perceive in the classic ant colony algorithm, i.e. inability to distinguish flows from different origins, inability of controlling distribution of vehicles along routes, inability to constrain flows or add link costs to routes. They incorporate pheromones and multi-colored ants to control distribution of traffic along links, and to distinguish between different traffic flows. Cesme and Furth [167] use self-organization to adapt traffic lights, including rules for extending green phases for arriving platoons (useful for long corridors) and a dynamic coordination to synchronize signals that are close to each other. They tested their methods in realistic scenarios with heterogeneous demands, showing that self-organization can reduce transit delay.

Wang, et al. [228] propose a control method that decomposes the network optimization problem into overlapped and interactive problems between "basic coordination units." Intersections use a forecast model, where not only the local performance is considered, but also that of the downstream intersections. Placzek [229] considers heterogeneous traffic flows and proposes a system where agents at intersections use models to predict the cost of control actions using the interval microscopic model, which combines cellular automata and interval arithmetic. After control actions are evaluated, the best is executed. Scenarios with road-side vehicle detectors and vehicular sensor networks are evaluated, considering heterogeneous traffic. The proposed mechanisms are shown to reduce delays compared with [160 and [185]. The advantage of Placzek method lies in the fact that it can adjust parameters in a heterogeneous fashion, while other methods use global parameters.

*4.2.2 Cellular Automata*

Cellular automata (CA) is another approach to model self-organizing systems that has been used to study traffic signal behavior and control since the 1950's [76]. Cellular automata models are described in greater detail in section 2.3 in the context of urban traffic modeling. Studies that have used CA with a focus on traffic signaling include [86]; [168]. Gershenson and Rosenblueth [157] use a CA model to examine traffic signaling and self-organization in relation to the problem of static-optimization of constantly changing traffic behavior. A macroscopic two-dimensional cellular automata model of an urban traffic signal control system, in which each intersection is regarded as a cell, and the flow pressure is treated as the state, is analyzed by Wei and colleagues [169], and Fouladvand and colleagues [80]. Wei and colleagues, using the BML model, simulate and analyze a two-dimensional traffic system controlled by traffic lights; Wu and colleagues [170] take a similar approach. Yang and colleagues [171], and Shoufeng and Ximin [172] examine travel and route guidance.

Shoufeng and Ximin combine a Hybrid Algorithm with a cellular automata simulation to calculate travel time and optimize signal settings. Paz et al. [173] integrate a genetic algorithm to search for a global optimum, and couple it with simulated annealing to get the best local optimum. While the results were quite similar to Simultaneous Perturbation, Stochastic Algorithm combination of a global and local optimum approach reduces the computational effort significantly. Kumar and Mitra [174], meanwhile, use a cellular automata model to examine what happens at an intersection when traffic signals malfunction. By modeling individual vehicles as agents, they were able to replicate the surprisingly well-organized traffic flow observed at a malfunctioning intersection in India; counter-intuitively, the behavior that normally causes jams induced traffic to flow smoothly.

In the Cell Transmission Model (CTM), a highway is partitioned into a series of cells, the densities of which are determined by vehicle concentration [150]. Flotterod [175] and Tampere and Immers [176] modify this idea by applying a Kalman filter to the model. The problem of multi-intersection control using the CTM and hierarchical control is addressed by Chen and colleagues [177], [178], who distinguish between feeding delay (the delay of vehicles entering a saturated cell), and nonfeeding delay (the delay of all other vehicles). Their proposed solution is formulated as a conflicted multi-objective optimization problem. In this type of problem, a single optimal solution may not exist. Instead, there may be a family of non-dominated or non-inferior solutions. These are the best solutions that can be obtained among the alternatives, such that increasing any one objective would degrade another. This is also known as a Pareto front, or Pareto optimal solution.

In the context of urban traffic modeling, it is interesting to note that bottlenecks create a situation similar to that described by self-organized criticality. Self-organized criticality (SoC) is a property of dynamic systems that have a critical point as an attractor. In other words, the system state naturally tends to move towards the attractor state, and that state, being at a critical point in the system, is near a phase transition, thus keeping the system in a dynamic, but self-organized situation. Thus, there is no need to precisely tune the system parameters to maintain this situation [179]. A famous example is the sand pile, in which, as sand is added to the pile, it will maintain a minimally stable state until it eventually collapses.

In summary, cellular automata is one of many techniques from statistical mechanics used to assess the micro- and macroscopic behavior of vehicles in traffic signaling. Many useful fundamental results have been obtained from these analyses. Current traffic engineering techniques rely on a combination of queuing theory and empirical techniques, and statistical mechanics approaches have not been fully integrated with these. As intelligent transportation becomes more mature, the need to incorporate these techniques will become increasingly significant.

*4.2.3 Neural Networks*

Neural networks are designed to observe and learn patterns using an architecture similar to that of brains, where multiple neurons are interconnected and learn to fire at different strengths in response to different environmental or internal stimuli, creating different impacts on the chemical and physical responses in the body. Similarly, an artificial neural network contains multiple interconnected neurons that learn from prior experience using algorithms, allowing the network to respond to previously unknown inputs. In effect, an artificial neural network builds its own regression model based on data used for its training, and then uses the model to predict outcome based on new input. In case of traffic, the model learns from the past traffic behavior and suggests an optimal timing sequence for traffic lights.

Ledoux [21] has used a back-propagation neural network to create a cooperation-based flow across multiple intersections. Such cooperation has many of the hallmarks of self-organization. The basic characteristic of such a neural network is that it can use data from past behavior to predict future behavior. Neural networks can be used to model traffic flow either at a single intersection or through multiple intersections. This enables observers to predict queue lengths and output flows one minute ahead of the decision point.

Nakatsuji & Kaku have also developed a self-organizing system using neural networks [20]. They trained a back-propagation neural network to predict traffic behavior. Input measures of cycle length and signal phase splits were used as inputs, with measures of effectiveness such as queue lengths or performance indexes as output. The authors make claims for self-organization in their model; however, it is more adaptive in nature insofar as each signal is able to adapt its behavior based on immediate conditions predicted by the neural network. There does not seem to be cooperation among the nodes in terms of information exchange.

*4.2.4 Utopian Solution*

Given the rapid improvements in sensing and communication that have allowed autonomous driving, in the future, knowledge of the location of cars, their routes, and velocities can be assumed. The cars are then akin to particles flowing through a network structure, and the goal is to avoid collision. For any given set of cars and complete knowledge of their paths, it should be possible to compute the optimum trajectory so that no collisions occur. For example, assume no traffic signals exist, and allow the cars to proceed until they reach their destinations. Then, note how many collisions would have occurred. These configuration estimates are typically based upon the requirement of maintaining a safe distance between cars, below which a collision is assumed. It is only at these collision points that traffic signals are required. Studies examining this idea include [180]; [181]; [182]; [230]

The question then becomes how to optimally apply signals. Should the cycle time of green to red be proportional to the traffic that actually arrives? Or should it be proportional to the traffic that is predicted to arrive? Prediction is feasible given that the shortest routes are assumed. One can further assume that vehicles will continue on a course that is as linear as possible with respect to the road topology. With these assumptions, the use of self-organizing traffic lights is not only feasible, but also very possible. Goel et al. [183] have developed an algorithm based on the predicted arrival of vehicles through communication among neighboring signals. Sensors located at the entrance to an artery detect a vehicle entering the artery, and relay the information to the next signal. Based on the congestion (computed by total number of cars in the artery) and posted speed limits, they can compute the arrival time to the next intersection. Based on the arrival times of vehicles at the intersection, an optimum signal time is computed. Goel et al. [184] have also developed a process for retrofitting existing legacy controllers with a microprocessor to support the self-organization algorithm. In contrast, Gershenson [36] relies on counting vehicles as they arrive at a traffic light, and does not require communication across traffic lights. Helbing and Lämmer [185] suggest a self-organizing algorithm where traffic conditions from the recent past are used to make future predictions to optimize the traffic signal timing, making them adaptive in real-time. This information is collected via sensors at individual intersections and fed into a microprocessor on the traffic signal to make optimal future signal timing decisions. Data collection and sensing are obviously keys to improving the efficiency and design of self-organizing algorithms. To identify the amount of data that needs to be collected, the impact of latency and precision of information on the overall performance needs to be measured. There are issues of cost, reliability, and privacy that also need to be addressed. We discuss data issues in depth in the next section.

# V. DRIVERS FOR SELF-ORGANIZATION: DATA COLLECTION & COMMUNICATION

## *5.1 Data Collection*

Traffic data collection is critical to signal timing and is often performed manually. Intelligent transport systems, on the other hand, continuously optimize signal timing and require high-quality real-time traffic data. A variety of sensors can be used for real-time data collection, the most common of which is the inductive loop detector [186]; [187]. The detector consists of one or more loops of wire embedded in the pavement and connected to a control box, activated by a signal ranging in frequency from 10 KHz to 200 KHz. When a vehicle passes over or rests on the loop, inductance is reduced, thus indicating the presence of a vehicle. These detectors work in most weather conditions, but can be unreliable and expensive. Ultrasonic sensors [188], microwave sensors [189], and infrared sensors [190] are also used to monitor traffic, though these have limitations as well. Ultrasound sensors are vulnerable to pedestrian and vehicular impact and have limited range (12m); infrared sensors perform poorly in noisy environments; and microwave sensors are unable to detect slow or stationary vehicles. Consequently, loop detectors remain the most popular technology in vehicle data monitoring.

Other new techniques for measuring traffic data include: radar [191]; [192]; cellular phones that record motorists' positions at check points [193]; Global Positioning Systems [194]; [195]; and RFID (radio-frequency identification) receivers which can be installed at strategic locations and are already used for toll collection [196]. Several other techniques are voluntary that require motorist cooperation. There is also a concern that some of these technologies violate personal privacy, and allow government agencies to monitor the behavior of private citizens.

The use of video surveillance is considered particularly pernicious by privacy advocates [197]; [198]; [199]. Nevertheless, vision-based systems are flexible and versatile in traffic monitoring applications; improvements in technology will likely continue to make these increasingly reliable and robust. Laser sensors used in autonomous vehicles are very effective for detecting vehicles, but their cost is still too high to be installed massively.

A robust tracking algorithm for crowded intersections has been developed by Kamijo [200]. This algorithm, based on the Spatio-Temporal Markov Random Field model, is robust against occlusion and clutter issues and can be modified to deal with illumination variation. This system is capable of generating traffic event statistics such as vehicle count, travel direction, velocity, and frequency paths.

With the advent of vehicular sensors there is a great potential for improving traffic efficiency. Sensor data can enable real-time control of traffic and improve the performance of existing adaptive systems by allowing a more precise adjustment of signal timing sequences.

*5.2 Communication*

There are two aspects of communication that are significant for self-organizing traffic systems: (1) sensor to traffic light, and (2) traffic light to traffic light. Sensor to traffic light has existed for decades, and is fairly robust. The key to self-organization is communication between traffic signals, which can be done using several different media such as fiber optic, radio signals, microwave signals, etc. depending on the topography, weather, and other physical constraints. While the technologies exist, there is a considerable cost in deploying such communication systems.

The vehicle to traffic light communication is rapidly evolving with considerable implications for traffic efficiency improvement. Audi already has pilot studies in several cities, including all of Berlin, where equipped cars get information from traffic lights and accelerate or reduce their speed depending on whether they will get a green light at the next intersection or not [201]. Okada and colleagues [202] propose the use of visible light wireless communication for vehicle-to-road communication at intersections. They recommend using a LED traffic light as the transmitter and an on-vehicle high-speed camera as the receiver. They also present an algorithm for detecting the transmission. The authors discuss a number of related factors, such as positioning of road signals and cameras, and how to achieve uninterrupted communication between traffic signals and moving vehicles.

Lee and colleagues [203] have suggested using a modified blackbody radiation model to capture the effect of ambient-light noise in the context of visible light communication. They study the impact of daylight on system performance and demonstrate that it is possible to obtain relatively accurate results with a relatively modest effort. The authors also introduce a new receiver structure with a selective combining technique that reduces the effect of background noise by 5 dBs.

Vehicle-to-vehicle communication is also novel, with even greater implications for traffic optimization using self-organization. Belinova [204] proposes the use of self-organization based on vehicle-to-vehicle communication. In her model, each vehicle is considered an independent intelligent entity, and a driver makes decisions based on information uniquely available to him/her from other vehicles and traffic signals. When the various reactions of individual drivers are combined, self-organizing processes may arise and contribute to the improvement of traffic, though the opposite is possible as well.

The Cluster-based Self-organizing Protocol, or CSP, was developed to organize and cluster vehicles [205]. This protocol assumes that each vehicle has the ability to communicate with a central server and exchange messages at intersections. Vehicles are clustered based on the information exchanged between them. This is somewhat analogous to a Mobile ad hoc network (MANET), in which one node acts as the base station and performs additional overhead tasks. In one MANET there are multiple service areas, with one sub-service area holding multiple mobile stations and one base station monitoring these.

An example of vehicle-to-vehicle communication for self-organizing highway traffic was developed by Kesting et al. [206]. A similar example was presented by Yang and Recker [207]. It is likely that this technology can also be extended to coordinate vehicles through intersections, as explored by Ilgin-Guler et al. [208]. Pehofer and Bettstetter [209] have explored self-organization based on ubiquitous wireless communication. They pose the fundamental question: What are the design paradigms for developing a self-organized network function? We attempt to answer that in the next section.

## VI. DESIGN OF SELF-ORGANIZING TRAFFIC SYSTEMS

Even though self-organized systems fall under the general rubric of complex systems, their most desirable attribute is that simple rules applied in a local neighborhood lead to sophisticated behavior globally. The key to this outcome, however, is finding the right set of rules. As previously noted, in nature many complex systems operate efficiently through the use of simple self-organization rules. While natural systems such as ant colonies, have precise, preordained, and predictable global outcomes, most man-made systems exhibit imprecise, unpredictable and chaotic behavior, and the difference is self-organization vs. central control.

Ashby introduced the concept of "requisite variety" which, loosely, posited that a variety of response mechanisms are necessary to deal with the variety of states a system can be in [210]. Considering urban systems as complex suggests that we cannot manage the complexity of large urban systems centrally, as a single central system does not have the variety of response mechanisms required to control the variety of all possible traffic states in a city. Solutions to traffic signal problems must distribute the coordination mechanism so that the complexity is allocated among and across the (self-organizing) controller modules.

Precisely controlled communication at the local level leads to desired behavior outcomes at the global level in natural systems. The desire to institute central control in artificial complex systems is inhibitive. In our efforts to control complex systems, we need to embrace their inherent qualities and behaviors in order to align them with their natural states, instead of forcing them under centralized control. Several artificial systems have effectively embraced self-organization, the most notable of which is the Internet, which uses self-organization by coordinating millions of relatively simple components to collectively create a virtual world exhibiting sophisticated behavior and properties. Social networking applications (*e.g*., Facebook, Twitter, and LinkedIn) also exhibit such behavior. Spontaneous eruption and scalability are hallmarks of these self-organized systems, which make them so potent.

A self-organized traffic system can similarly be designed by exploiting the natural order to optimize flow through the entire traffic grid by using simple rules that govern a single signal's behavior while enabling global synchronization. Four paradigms are germane to the development of self-organized systems: (1) designing local interactions that achieve global properties; (2) exploiting implicit coordination; (3) minimizing the maintained state; and (4) designing protocols that adapt to change.

Associating complex systems with traffic control and developing the proper rules for interaction and behavior can be a difficult undertaking. Traditional optimization techniques have a role in the design of complex systems. We propose implementing a self-organizing traffic model using classical optimization techniques for local optimization at each intersection, and coupling with adjacent signals through offsets that can be varied through simulated annealing or some other stochastic method. Over time, the offsets can be synchronized pairwise across the entire grid, resulting in global synchronization. The amount of information (*e.g*. sensors, statistical data flows) used in the optimization also influences the performance of such algorithms. Consequently, local optimization algorithms, rules of coupling across intersections, and the information available for optimization, all need to be investigated while developing self-organized traffic models.

There are fundamental questions that need to be answered based on any model that we develop. For instance, organisms that self-organize successfully use a limited number of attributes in decision-making. What is the minimal set of rules that can be applied to efficient systems? This leads us to the notion of complexity. Our goal should be not only to determine the minimal subset of rules, but also to understand the limits of self-organization. As the number of attributes increases, will self-organization still be feasible? For instance, the free market economy is a self-organizing system with a basic set of rules that all participants must follow. It generally works efficiently; however, its rules occasionally need to be tweaked as participants capitalize on systematic inefficiencies and disrupt its natural harmony. Such possible flaws in the rules of self-organization for traffic control could lead to dangerous traffic conditions and need to be identified. All of these are relevant research problems that need to be addressed.

Self-organizing traffic signals promise to revolutionize current traffic control mechanisms and have the potential for results beyond our expectations. For instance, Audi is already testing vehicular speed controllers that will adjust speed based on traffic congestion and signal timing, such that the car gets a green light when it reaches an intersection [201]. There are broader implications to this work; urban planning may need to be reviewed and traffic rules modified.

There are other innovative ways to think about traffic optimization using techniques that become feasible as sensing and communication technologies develop further. For instance, the network structure formed by the physical and logical connections of traffic intersection signals can also be studied using graphs, such that each node is a signal and each link is an artery. Traffic behavior can also be examined in a manner akin to communication network analysis. Scheduling in traffic networks differs insofar as cars do not have destination addresses; their precise routes cannot be known with certainty. In addition, there is a wavelike synchronization issue in which traffic that starts at the right time and flows at the right velocity can "catch" all the green lights. In fact, if individual cars could be scheduled to leave at a scheduled time and maintain a constant speed, it would be possible to minimize both delay and the very need for traffic lights by having the flows become "orthogonal" to one another, in other words, able to flow through one another without collision. This seems utopian right now, but with collision avoidance systems and vehicular sensors this may become a reality sooner than we realize.

An intersection scheduling model based on individual vehicle arrivals has been developed by Wu and colleagues [211]. They schedule the use of the intersection by different vehicles based on expected arrival time. This approach assumes that there are no traffic lights at the intersection, but rather an intelligent device that determines the schedule of intersection use for vehicles based on location, velocity, urgency, etc. Decomposition and backtracking logic (at each time unit) are used to estimate which lane to release to minimize waiting time. Vehicles negotiate their access time through an intelligent device embedded in the intersection. The scheduling mechanism is based on the space between two successive vehicles in the conflicting traffic lane, and minimum space between successive vehicles. The authors show that negotiation between vehicles by means of well-adapted approaches can be effective in improving traffic control at a simple intersection. A central server manages and monitors the traffic in each intersection. This of course assumes that all vehicles follow the command of the central controller, reducing considerably the complexity of the traffic flow and facilitating its prediction. Thus, while we still have independently interacting vehicles, we must build traffic light systems that match the complexity of the traffic flow. Self-organization offers one way of achieving this.

# VII. CONCLUSIONS

There are many factors which contribute to traffic flow, including driver behavior, urban topology, transit rules, driving conventions, infrastructure, and demand. Here we have focused on one factor only: traffic signals. As we have argued in the preceding analysis, traffic signal design is a multivariate optimization problem involving numerous decision variables such as cycle lengths, split times and offsets. It is also a distributed optimization problem in which the behavior of each traffic light can be controlled independently. Finally, it is a coupled optimization problem in which changes in the timing patterns of one signal will influence the performance of other signals in the grid. Ideally, we would like a closed-form solution to the problem such that all the traffic lights are optimized simultaneously (cycle lengths and timing splits are coordinated) so as to minimize the aggregate delay across the entire grid. However, the problem is far too complex to obtain a closed-form solution. Moreover, the specific traffic conditions change constantly, *i.e.* traffic is a non-stationary problem, and so even if we were able to find an optimal solution, this would be obsolete in seconds. In general, the problem can be approached iteratively, whereby solutions are created and evaluated successively to improve the design progressively. Stochastic, gradient-based, and heuristic techniques have all been used for traffic signal optimization. Both real-time and a priori approaches can be used with their own unique pros and cons. The unpredictability of traffic patterns undermines the effectiveness of a priori models, while high latency and lack of scalability have a negative impact on real-time models.

We can optimize complex systems by decoupling the problem naturally using the principles of self-organization. This not only makes the solutions efficient, but also adaptable, as discussed in previous sections. The traffic signal optimization problem is a classic case of such a complex system in which a network of heterogeneous components interacts nonlinearly, giving rise to emergent behavior. Agents interact with each other and make independent local decisions based on input from the environment.

While the optimality of results cannot be gauranteed, they are more responsive to local conditions, and are closer to the theoretical optimum compared to existing techniques. Preliminary attempts have been made to address the problem of traffic signal synchronization using self-organizing approaches. We believe that this is a promising area of research that remains largely unexplored. A variety of optimization algorithms and traffic condition data sets could be tested at the local level to investigate the impact of various datasets, and to test the robustness of models in a variety of weather conditions. Benchmarks and optimality measures are required for comparing different proposals in simulations. In this direction, Zhang et al. [231] have used macroscopic fundamental diagrams to compare different traffic light coordination mechanisms, showing that self-organizing methods achieve overall better performance and a higher network capacity than centralized or adaptive control mechanims.

In practice, the suitability of such techniques depends on the quality of real-time traffic data in the local signal grid, which comes at a cost. With advances in sensor and communication technology, the availability and quality of data will continue to improve. However, there is no doubt that communications and sensors have advanced sufficiently to allow us to move such a system from conception to reality.

There are several unanswered questions about the design of such systems that will need to be addressed, and are potent topics of research. The research on complex systems in this area is still rudimentary, and needs to be examined further. Also, with advances in vehicular sensors, we may be able to create more innovative models for traffic optimization by controlling the behavior of vehicles, rather than just traffic lights. The work of Berger and Rumpe [212], for example, shows that the development of autonomous vehicles is already moving in this direction. Given the trend towards interconnectivity, it is only a matter of time before vehicles communicate with one another and with infrastructure. With greater research, lessons from natural self organized systems could inform complex transportation systems towards greater precision of controlled communication at the local level leading to desired behavior outcomes at the global level.

ACKNOWLEDGEMENTS

"The author(s) declare(s) that there is no conflict of interest regarding the publication of this paper."

*Computational Engineering in Systems Applications*, 2006.

[172] L. Shoufeng e L. Ximin, "Based on Hybrid Genetic Algorithm and Cellular Automata Combined Traffic Signal Control and Route Guidance," em *Chinese Control Conference*, Changsa, 2007.

[173] A. Paz, V. Molano, E. Martinez, C. Gaviria e C. Arteaga, "Calibration of traffic flow models using a memetic algorithm," *Transportation Research Part C: Emerging Technologies,* vol. 55, pp. 432-443, 2015.

[174] S. Kumar e S. Mitra, "Self-organizing Traffic at a Malfunctioning Intersection," *Journal of Artificial Societies and Social Simulation,* vol. 9, nº 4, 2006.

[175] G. Flotterod, "Some practical extensions to the cell transmission model," em *IEEE Intelligent Transportation Systems*, 2005.

[176] C. M. Tampere e L. H. Immers, "An Extended Kalman Filter Application for Traffic State Estimation Using CTM with Implicit Mode Switching and Dynamic Parameters," em *IEEE Intelligent Transportation Systems Conference*, 2007.

[177] J. Chen, L. Xu, X. Yang e C. Yuan, "A hierarchy control algorithm and its application in urban arterial control," em *IEEE Intelligent Transportation Systems*, 2007a.

[178] J. Chen, L. Xu, X. Yang e C. Yuan, "Oversaturated adjacent intersections control based on multi-objective compatible control algorithm," em *IEEE Intelligent Transportation Systems*, 2007b.

[179] P. Bak, C. Tang e K. Wiesenfeld, "Self-organizaed Criticality: An Explanation of the 1/f Noise," *Physical Review Letters,* vol. 59, nº 4, pp. 381-4, 1987.

[180] C. Gershenson, "Control de tráfico con agentes: CRASH," em *Memorias XI Congreso Nacional ANIEI*, Xalapa, 1998.

[181] K. Dresner e P. Stone, "Multiagent Traffic Management: A Reservation-Based Intersection Control Mechanism," em *The Third International Joint Conference on Autonomous Agents and Multiagent Systems*, 2004.

[182] M. Vasirani e S. Ossowski, "A computational market for distributed control of urban road traffic systems," *IEEE Transactions on Intelligent Transportation Systems ,* vol. 12, nº 2, pp. 313-321, 2011.

[183] S. Goel, S. F. Bush e K. Ravindranathan, "Self-organization of traffic lights for minimizing vehicle delay," em *Proceedings of the 3rd International Conference on Connected Vehicles and Expo (ICCVE), IEEE*, Vienna, Austria, 2014.

[184] S. Goel, E. Dincelli, A. Parker e E. Sprissler, "Enabling a self-organized traffic system in existing legacy hardware," em *Proceedings of the 3rd International Conference on Connected Vehicles and*
48

**BIOGRAPHIES**

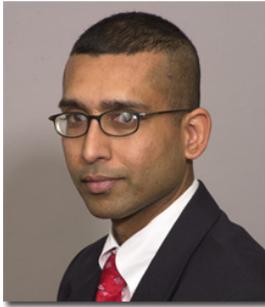

Sanjay Goel is an Associate Professor in the School of Business and the Director of Research at the NYS Center for Information Forensics and Assurance at UAlbany. He represents UAlbany in the Capital Region Cyber Crime Partnership. Dr. Goel received his Ph.D. in Mechanical Engineering from RPI. His research includes information security, risk analysis, security policies, information classification, cyber warfare and self-organization in complex systems. His latest research on self-organizing systems includes traffic light coordination, nano-bio computing and social networks. He and his team have worked with CSCIC in developing the information classification policy for New York. He is currently leading an effort launched by IEEE Communications Society and the IEEE Standards Association to create a vision for the Smart Grid future 15 years ahead. He won the promising Inventor's Award in 2005 from the SUNY Research Foundation. In 2006, he was awarded the SUNY Chancellor's Award for Excellence in Teaching, the UAlbany Excellence in Teaching Award, and the Graduate Student Organization Award for Faculty Mentoring. He was named an AT&T Industrial Ecology Faculty Fellow for 2009-2010. He has received grants and funding from NIJ, NSF, UTRC, NYSERDA, US Department of Education, and CSCIC.

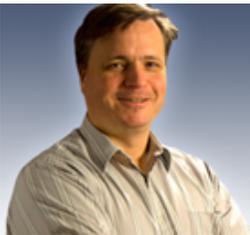

Stephen F. Bush is a researcher in Algorithmic Communications Network Theory at the GE Global Research Center. Dr. Bush was presented with a Gold Cup Trophy Award from DARPA for his work in fault tolerant networking. Stephen F. Bush received the B.S. degree in electrical and computer engineering from Carnegie Mellon University, M.S. degree in computer science from Cleveland State University, and Ph.D. degree from the University of Kansas. He is currently a researcher at General Electric Global Research, Niskayuna, NY. He is the author of Nanoscale Communication Networks (Norwood, MA: Artech House, 2010). He coauthored a book on active network management, titled Active Networks and Active Network Management: A Proactive Management Framework (New York, NY: Kluwer Academic/Plenum Publishers, 2001). He is an internationally recognized researcher in Active Networking and Algorithmic Communications Networking Theory with over 75 peer-reviewed publications. Dr. Bush is the past chair of the IEEE Emerging Technical Subcommittee on Nanoscale, Molecular, and Quantum Networking and currently chair for the IEEE 1906.1 standards working group on nanoscale communication networks. Dr. Bush is an IEEE Distinguished Lecturer on the smart grid and nanoscale communication networks. He is also on the steering committee for the IEEE Smart Grid Vision Project.

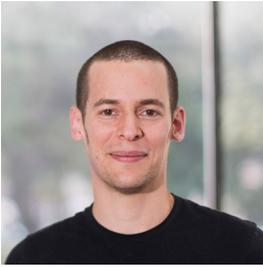

Carlos Gershenson is a full time research professor at the Instituto de Investigaciones en Matemáticas Aplicadas y en Sistemas at the Universidad Nacional Autónoma de México (UNAM), where he leads the Self-organizing Systems Lab. He is also affiliated researcher and member of the directive council at the Center for Complexity Sciences at UNAM. He is Visiting Professor at the Massachussetts Institute of Technology and at Northeastern University. He has a wide variety of academic interests, including self-organizing systems, artificial life, evolution, complexity, cognition, artificial societies, and philosophy. He is Editor-in-Chief of Complexity Digest (http://comdig.unam.mx) and Complexity-at-large editor for the journal Complexity.